\documentstyle[twoside,fleqn,espcrc2,epsf]{article}

\newcommand{\AmS}{{\protect\the\textfont2
  A\kern-.1667em\lower.5ex\hbox{M}\kern-.125emS}}

\title{  Visualization of topological structure and chiral condensate 
         \thanks{Supported in part by FWF under Contract No. P11456} 
}

\author{Markus Feurstein, Harald Markum and Stefan Thurner \\
\vspace{3mm}
Institut f\"{u}r Kernphysik, TU Wien,
         Wiedner Hauptstra\ss e 8-10, A-1040 Vienna, Austria\\  
} 
       
\begin{document}

\begin{abstract}
We perform a mutual analysis of the topological and chiral vacuum structure 
of four-dimensional QCD on the lattice at finite temperature. 
We demonstrate that  at the places  
where instantons are present, amplified production of 
quark condensate takes place. It turns out  for full QCD that the clusters of 
nontrivial chiral condensate have a size of about 0.4 fm 
corresponding  to the instanton sizes  in the same configurations. 
\end{abstract}

\maketitle


It is well known that the zero eigenvalues of the fermionic matrix 
are related to the global topological charge $Q$ of a gauge field configuration 
via the Atiyah-Singer index theorem.
It is believed that the instantons as carriers of the topological charge  
might play a crucial role in understanding 
the confinement mechanism of four-dimensional QCD, if one assumes that 
they  form a so-called instanton liquid \cite{SHU88}.
Recently, it was demonstrated that monopole currents appear preferably 
in the regions of non-vanishing topological charge density \cite{wir,andere}.
It has been conjectured that both instantons and monopoles are related 
to chiral symmetry breaking \cite{SHU88,MIA95}. This idea is further 
supported by the following results of a direct investigation of the local 
correlations  
of the quark  condensate and the topological charge density. 

For the implementation of the topological charge on a Euclidian lattice
we restrict ourselves to the so-called field theoretic definitions which
approximate the topological charge density in the continuum,
$
q(x)=\frac{g^{2}}{32\pi^{2}} \epsilon^{\mu\nu\rho\sigma}
\ \mbox{\rm Tr} \ \Big ( F_{\mu\nu}(x) F_{\rho\sigma}(x) \Big ) \ .
$
We used the plaquette and the hypercube prescription.
To get rid of  quantum fluctuations and 
renormalization constants,
we employed the Cabbibo-Marinari cooling method.
Mathematically and numerically  
the local chiral condensate $\bar \psi \psi (x)$ 
is a diagonal element of the inverse of the fermionic matrix
of the QCD action. 
We compute correlation functions between two observables 
${\cal O}_1(x)$ and ${\cal O}_2(y)$ 
\begin{equation}
\label{correlations}
g(y-x)=\langle {\cal O}_1(x) {\cal O}_2(y) \rangle - 
       \langle {\cal O}_1\rangle \langle {\cal O}_2\rangle
\end{equation}
and normalize them  to the smallest lattice 
separation $d_{\rm min}$,  $ c(y-x)=g(y-x)/g(d_{\rm min})$. 
Since  topological objects with opposite sign are equally distributed,
we correlate the  
quark-antiquark density  with the square of the topological charge density.
\begin{figure}[t]
\vspace{-1mm}
\epsfxsize=7.5cm\epsffile{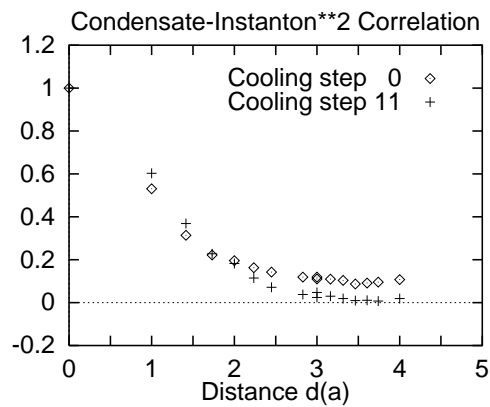 }  
\vspace{-10mm}
\caption{ 
Correlation function of the quark-antiquark density  
and the topological charge density  for 0 and 11 cooling steps. 
The  correlations extend over two lattice spacings and indicate local 
coexistence  of the quark  condensate and topological objects.
}
\vspace{-8mm}
\label{corr}
\end{figure}
\begin{figure}
\vspace{-17mm}
\begin{tabular}{c}
5 Cooling steps\\
\epsfxsize=6.5cm\epsffile{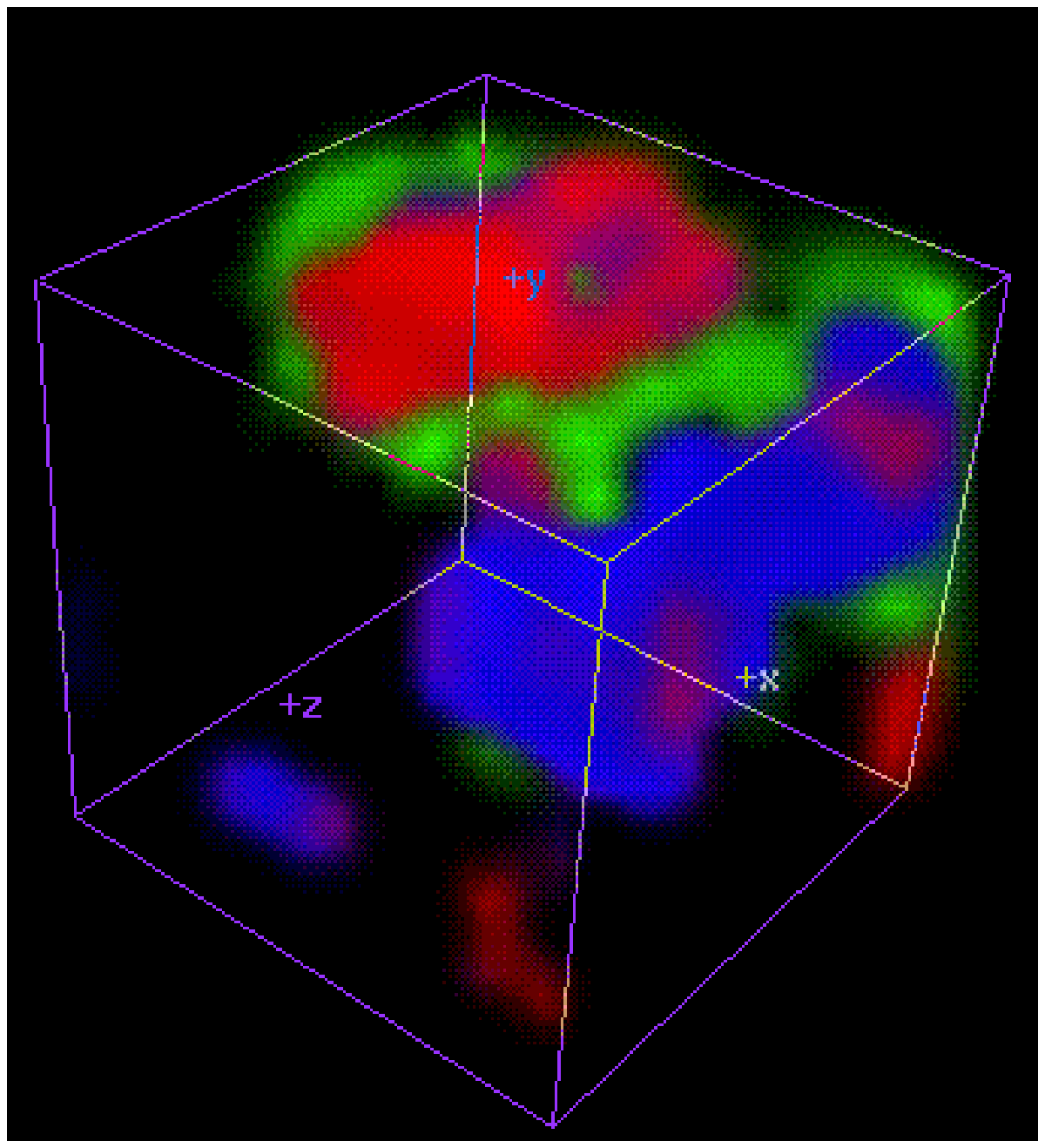}\\
 \\
10 Cooling steps\\
\epsfxsize=6.5cm\epsffile{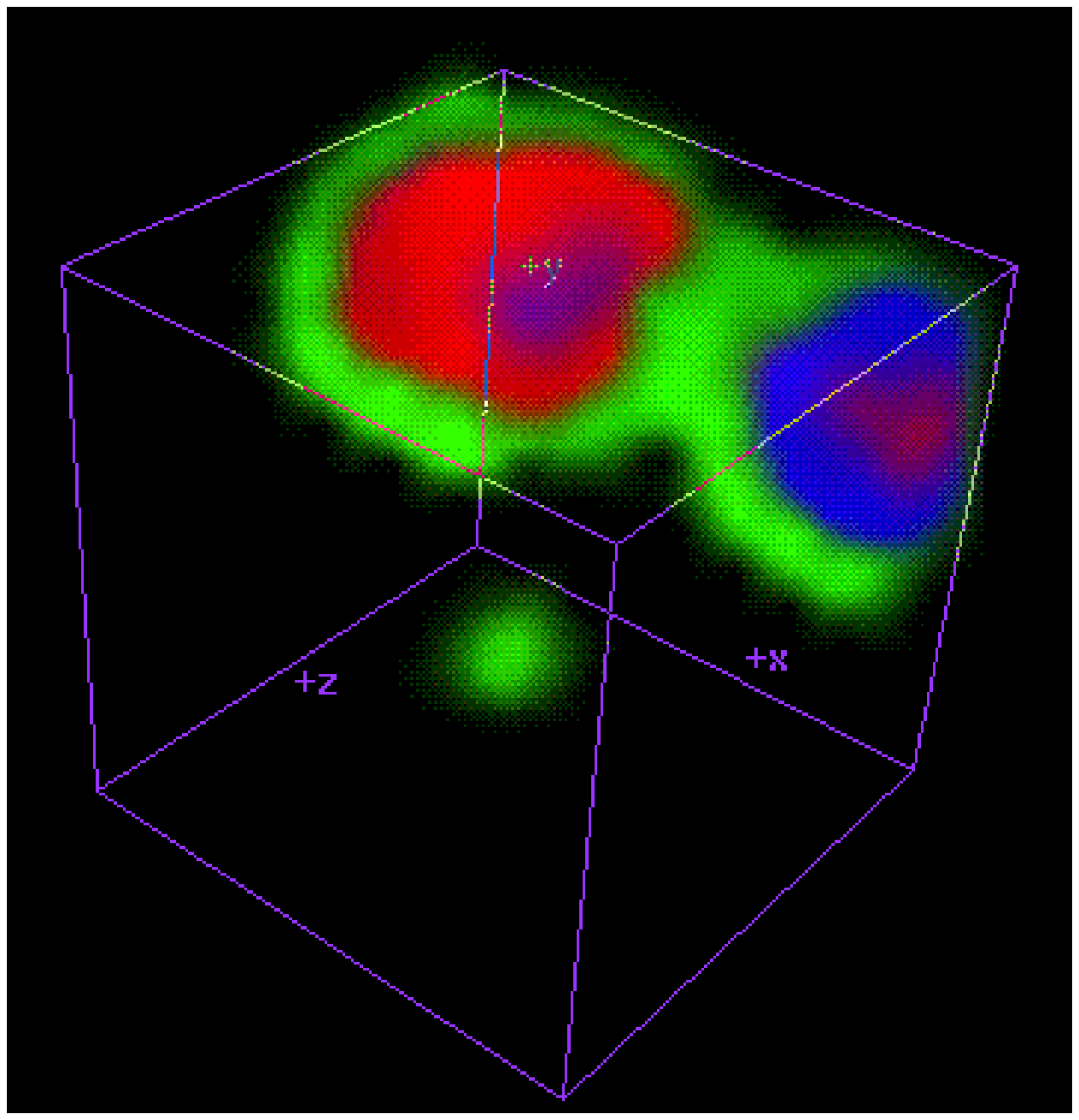 }  \\
\end{tabular}
\vspace{-3mm}
\caption{
Cooling history for a time slice of a single gauge field
configuration of $SU(3)$ theory with dynamical quarks.
The dark  and medium grey shades  represent the positive and negative 
topological charge density respectively; 
the surrounding light grey tone  the density of the quark  condensate. 
It turns out that the quark condensate 
takes a non-vanishing value at the positions of the instantons.  }
\label{hist}
\vspace{-10mm}
\end{figure}
Our simulations were performed for full $SU(3)$ QCD on an 
$8^{3} \times 4$ lattice with
periodic boundary conditions.  Applying a standard Metropolis algorithm 
has the advantage that tunneling between sectors of different topological 
charges occurs at reasonable rates. 
Dynamical quarks in Kogut-Susskind discretization   
with 3 flavors of degenerate  mass $m=0.1$ were taken into account using the 
pseudofermionic method. 
We performed runs  in the confinement phase at $\beta=5.2$. 
Measurements were taken on 1000 configurations separated 
by 50 sweeps.

Figure~\ref{corr} shows results for the correlation functions 
of Eq.~(\ref{correlations}) with 
${\cal O}_1=\bar \psi \psi(x)$ and $ {\cal O}_2=q^2(y)$. 
The $\bar \psi \psi q^2$-correlation function  exhibits an extension  
of more than two lattice spacings, indicating nontrivial 
correlations. 
To gain information about the correlation lengths exponential 
fits to the tails of the correlation 
functions were performed. They show that an increasing number of cooling steps
yields shorter  correlation lengths. The corresponding 
screening masses are $\zeta=0.59$ and $\zeta=1.56$ 
in inverse lattice units for 0 and 11 cooling steps, respectively. They 
have to be interpreted as effective masses and reflect the effective gluon 
exchange and the vacuum structure of QCD.

It is assumed that the size $\rho$ 
of a t'Hooft instanton $q_\rho(x)\sim  \rho^4 \, (x^2+\rho^2)^{-4}$ 
centered around the origin enters also 
into the associated distribution of the chiral condensate 
$\bar \psi \psi_\rho (x) \sim \rho^2 \, (x^2+\rho^2)^{-3}$ 
\cite{shurev}. 
To estimate $\rho$ we fitted a convolution of the functional form 
$f(x)=\int \bar\psi\psi_\rho(t) q_\rho^2(x-t)\, dt$ 
to the data points. 
This was evaluated after 11 cooling steps where the configurations are 
reasonably dilute. Our fit yields $\rho(\bar\psi\psi q^2) 
=1.8$ in lattice spacings. 
To check consistency we extracted from the $qq$-correlation 
a value of $\rho(q q)=2.05$ which is in good agreement. 

We now visualize densities  of the quark  condensate and
topological quantities from individual gauge fields. The topological content
of typical individual gauge field configurations was already studied for pure
$SU(2)$ and $SU(3)$ theory. We found that at the local regions
of clusters of topological charge density, which are identified with
instantons, there are monopole trajectories looping around in almost all 
cases \cite{wir}. 

\begin{figure}
\vspace{-15mm}
\begin{tabular}{c}
\epsfxsize=7.5cm\epsffile{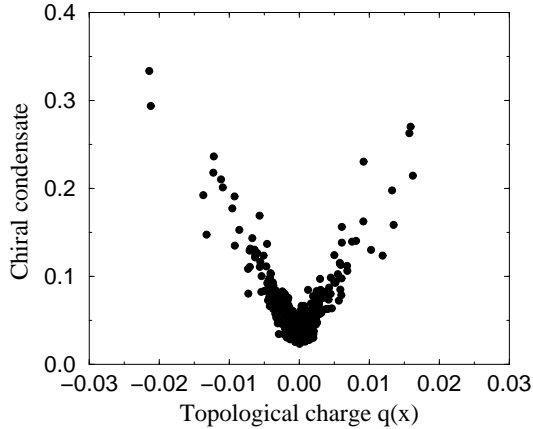}\\
\end{tabular}
\vspace{-10mm}
\caption{
Scatter plot of $\bar \psi\psi (x)$ against $q(x)$ in the volume 
of the same gauge field configuration with 10 cooling 
steps as in Fig.~2. It reflects the separation into an instanton 
and antiinstanton. 
}
\vspace{-7mm}
\label{scatt}
\end{figure}
In Fig.~\ref{hist} a time slice of a typical configuration from $SU(3)$ theory
with dynamical quarks on the $8^{3} \times 4$ lattice 
in the confinement phase is shown.
We display the positive instanton density by dark grey shades 
and the negative density by a medium grey-tone if the absolute value
$|q(x)| > 0.003$.
The quark-antiquark density  is indicated by a light grey shade whenever 
a threshold for $\bar \psi\psi (x) > 0.066$ is exceeded.
By analyzing dozens of gluon and quark field configurations we found the
following results.
The topological charge is covered by quantum fluctuations and
becomes visible by cooling of the gauge fields. For
0 cooling steps no structure can be seen in $q(x)$, $\bar \psi\psi (x)$
or the monopole currents, which does not mean the absence of correlations
between them. After 5 cooling steps clusters of
 nonzero topological charge density and quark  condensate are resolved.
This particular configuration possesses a cluster with a positive and a negative
topological charge  corresponding to an instanton and an antiinstanton,
respectively. For more than 10 cooling steps both topological charge and
chiral condensate begin to die out and eventually vanish.
Combining the above finding of  Fig.~1 showing that the correlation 
functions between $\bar \psi\psi (x)$ and $q^2(y)$  
are not very sensitive to cooling together with the cooling history of the 
3D images in Fig.~2, we conclude that instantons go hand in hand 
with clusters of 
$\bar \psi\psi (x)\neq 0$ also in the uncooled QCD vacuum \cite{HANDS}.

Figure~2 has demonstrated that the quark-antiquark density   
attains its maximum
values at the same positions where the extreme values of the topological
charge density are situated. 
This behavior is further substantiated in Fig.~\ref{scatt} where the
$\bar \psi\psi (x)$-values are plotted against $q(x)$ for all
points $x$ in the same configuration at 10 cooling steps. 
At first sight a linear relationship between the absolute value 
of the topological charge density  and the 
virtual quark density is suggested.


In summary, our calculations of correlation functions
between topological charge and the quark  condensate yield an
extension of about two lattice spacings. The correlations suggest that the
local chiral condensate takes a non-vanishing value predominantly in the regions  of
instantons and monopole loops.
It was well known before that the chiral condensate is related to
the topological charge and topological susceptibility. 
The visualization exhibited that the distribution of the ``chiral condensate''
concentrates around areas with enhanced topological activity 
(instantons, monopoles).
We demonstrated that exactly at those places in Euclidean space-time, 
where tunneling between the vacua occurs, amplified production of 
quark condensate takes place. 
It must be emphasized that this represents the situation on a finite lattice with 
finite quark mass without the extrapolation  to the thermodynamic and chiral limit. 
We found for full $SU(3)$ QCD with dynamical quarks that the clusters
of non-vanishing  quark condensate have a size of about $0.4$ fm,
which corresponds to the instanton sizes
observed in the same configurations. 
Visualization of quark and gluon fields might be especially useful to decide 
if the instanton-liquid model is realized in nature and 
if instanton-antiinstanton pairs appear in the deconfined phase.
It might also help to clarify the question of the existence of a 
disoriented chiral condensate with consequences for heavy-ion 
experiments.
\vspace{-4mm}
%


\end{document}